# Blind Known Interference Cancellation

Shengli Zhang*, Soung Chang Liew[§], Hui Wang*
*Communication Engineering Department, Shenzhen University, Shenzhen, China
[§] Information Engineering Department, Chinese University of Hong Kong, Hong Kong, China

**Abstract:**

This paper investigates interference-cancellation schemes at the receiver, in which the original data of the interference is known *a priori*. Such *a priori* knowledge is common in wireless relay networks. For example, a transmitting relay could be relaying data that was previously transmitted by a node, in which case the interference received by the node now is actually self information. Besides the case of self information, the node could also have overheard or received the interference data in a prior transmission by another node. Directly removing the known interference requires accurate estimate of the interference channel, which may be difficult in many situations. In this paper, we propose a novel scheme, Blind Known Interference Cancellation (BKIC), to cancel known interference without interference channel information. BKIC consists of two steps. The first step combines adjacent symbols to cancel the interference, exploiting the fact that the channel coefficients are almost the same between successive symbols. After such interference cancellation, however, the signal of interest is also distorted. The second step recovers the signal of interest amidst the distortion. We propose two algorithms for the critical second steps. The first algorithm (BKIC-S) is based on the principle of smoothing. It is simple and has near optimal performance in the slow fading scenario. The second algorithm (BKIC-RBP) is based on the principle of real-valued belief propagation. It can achieve MAP-optimal performance with fast convergence, and has near optimal performance even in the fast fading scenario. Both BKIC schemes outperform the traditional self-interference cancellation schemes with perfect initial channel information by a large margin, while having lower complexities.

## I. Introduction

The use of relay in wireless networks is attracting increasing attention [1, 2] because of the many advantages it brings, such as improved connectivity and reduced power consumption. Many multi-hop relay standards, including 802.16j, 802.11s, are being developed.

In wireless relay networks, a node may receive a target signal superimposed with interferences. However, in many scenarios, the receiver actually knows the data contained in the interference [3], either because the interference was signal previously received by the node, or the interference is self-information previously transmitted by the node and now transmitted by the relay superimposed with signal from another node. One example of known interference is when physical-layer network coding over infinite field [4] (e.g., analog network coding [5]) is used in a two-way relay channel, as shown in Fig. 1a. Another example is a linear-chain one-way relay network [6], as shown in Fig. 1b. Many other scenarios of known interference can be found in [3].

The method to deal with known interference is straightforward in theory. The receiver first estimates the channel coefficient of the interference signal and then removes the known interference from the target signal [3]. We refer to this scheme as traditional KIC (Known Interference Cancellation) in this paper. In practice, however, the scheme does not perform well when the channel estimation is inaccurate.

Accurate channel estimation is a non-trivial problem even in the absence of interference. This is the reason why non-coherent detection schemes that do not require channel information are still widely studied and used in wireless communication systems [7]. In the presence of interference, we face additional challenges. First, channel estimation is more difficult because the training sequences are corrupted by the superposition of two signals. Estimation of the channel coefficients of the targeted packet and the interfering packet can be complex. For example, when the two training sequences are the same and they overlap with each other, we can only obtain the summation of the two channel coefficients but not their individual values. Second, when the power of the interfering packet is much larger than that of the target packet, a tiny estimation error on the interference channel may cause the interference cancellation process to leave behind a relatively large residual interference with respect to the power of the target packet. Third, it may be impossible to estimate the channel accurately [8] in mobile environment with fast fading. The channel estimated from the training sequence may have changed by the time the data is received.

We note that the above difficulties apply regardless of whether the interference is known or unknown. This is because for both cases, the training sequences are presumed known in order that channel estimation can be enabled. Basically, we need to estimate the channels of two superimposed packets. There has been some work trying to tackle this problem. For example, in physical-layer network coding [9, 10], the channels of two superimposed packets need to be estimated. To deal with the channel estimation problem, the physical-layer network coding implementation in [11] uses orthogonal sequences for the two packets. The analog network coding scheme in [5], on the other hand, attaches the training sequence to both the front end and back end of a packet, and time the transmissions of the two packets so that one of them has interference-free front end and the other one has interference-free back end. These schemes use new frame designs and are not compatible with legacy wireless systems. Ref. [12] uses an optimization scheme to estimate the channels with two specially designed training sequences. The estimation accuracy is much poorer than that in single-channel estimation. Non-coherent ANC schemes that avoid channel estimation at the end nodes have also been studied [13, 14]. However, non-coherent schemes suffer from SNR degradation of about 3dB compared with the coherent schemes. By contrast, the blind known interference cancellation (BKIC) schemes proposed in this paper can obtain near-perfect performance – specifically, performance close to that of a point-to-point communication link without interference – while avoiding estimation of the interference channel.

BKIC has three advantages over the traditional methods: 1) better performance; 2) no need for interference channel estimation; and 3) compatibility with legacy systems. The principle on which BKIC operates is based on the observation that the wireless channel typically remains almost unchanged between adjacent symbols [8]. BKIC uses the interference in one symbol to cancel the interference in its adjacent symbol. For example, if the interference channel is *h* and the interference symbol is 1, then the interference in the current symbol is *h*. If the interference data in the adjacent symbol is -1, then the corresponding known interference is approximately –*h*. The interference can be cancelled with each other if we combine the two symbols. Such adjacent-symbol combination, however, may result in distortion of the target signal. Thus, a key issue is how to remove such distortion as the next step. We propose two schemes, smoothing (BKIC-S) and real-valued belief propagation (BKIC-RBP), to equalize the resulting distortion.

This paper considers BKIC schemes for both flat fading channel and frequency selective channel. We show that the performance of our schemes is almost the same as that of a pure coherent point-to-point channel without interference (note: this is a theoretical upper bound for our system). Besides its excellent performance, our schemes are also attractive because of their low complexity and compatibility with legacy systems. Specifically, our schemes do not require special changes to the frame structure or the operation of the transmitter. A salient feature of our schemes is that they could be realized by an add-on module inserted into the signal processing path of the receiver without requiring complicate modifications to the existing module.

The remainder of this paper is organized as follows. In section II, we present the system model and architecture of BKIC. Section III explains BKIC under the flat fading channel assumption, and Section IV extends the discussion to the frequency selective fading channels. In Section V, we analyze the performance. We validate and supplement the analytical results with numerical simulation in Section VI. Finally, Section VII concludes this paper.

## II. System Model and Architecture

Known interference is common in many wireless networks. TWRC with analog network coding [5] in Fig. 1a, and one-way relay chain network in Fig. 1b, are two examples with known interference. In this section, we present the general mathematical formulation for known interference systems. For a focus, consider the chain network in Fig. 1b. The source node $S$ transmits to the destination node $D$ through two relay nodes, $R_1$ and $R_2$, and there are no cross-hop transmissions.

For simplicity, we assume one dimensional $q$-ary ASK modulation at all the nodes; our method can be easily extended to other modulations, including two-dimensional modulations. As shown in Fig. 1b, in time slot 1, $S$ transmits a packet to relay $R_1$; in time slot 2, $R_1$ forwards it to the second relay $R_2$; and in time slot 3, $R_2$ forwards it to destination $D$, while node $S$ sends a new packet to $R_1$ at the same time. The transmissions in time slots 2 and 3 are repeated for the delivery of successive packets from $S$ to $D$.

Note that node $R_1$ receives a superposition of the two packets, one from $S$ and one from $R_2$. With the

assumption of symbol level synchronization [15], the $k$-th received symbol at $R_1$ can be expressed as

$$r(k) = \sum_{l=0}^{L_x} h_x(k,l)x(k-l) + \sum_{l=0}^{L_I} h(k,l)I(k-l) + n(k) \qquad (1)$$

where $x(k), I(k) \in A = \{-q+1, -q+3, \cdots q-3, q-1\}$ is the $k$-th target symbol sent from $S$ and $I(k)$ is the $k$-th interfering symbol sent from $R_2$; $n(k)$ is the zero mean Gaussian noise with variance $\sigma^2$; $h(k,l)$ and $h_x(k,l)$ are respectively the $l$-th tap channel coefficients from $R_2$ and $S$ to $R_1$ for their $k$-th symbols; $L_x$ and $L_I$ are the respective maximum tap delays of the two signals. The transmit powers, and the effects of transmit and receive pulse shapes, are combined into the channel coefficients, $h(k,l)$ and $h_x(k,l)$. According to the WSSUS model of Bello [16], all the channel taps are independent of each other; for each tap, the channel variation satisfies $E\{h(k,l)h(k-1,l)\} = J_0(2\pi f_{max}\tau)$, where $J_0(2\pi f_{max}\tau)$ denotes the zero-th order Bessel function of the first kind, $f_{max}$ is the maximum Doppler frequency, and $\tau$ is the symbol duration. Hereafter, we use the bold $x/I$ letter to denote the corresponding vector of the whole packet.

The system formulation for analog network coding in two-way relay channel is the same as in (1) and the details are omitted here.

**System Architecture:**

In this paper, we propose a blind known interference cancellation scheme which can cancel the interference in (1) and transform $r(k)$ to the signal of interest, $\sum_{l=0}^{L_x} h_x(k,l)x(k-l) + n(k)$, plus a small noise as

$$\begin{aligned} z(k) &= \sum_{l=0}^{L_x} h_x(k,l)x(k-l) + n(k) + w(k) \\ &= x'(k) + n(k) + w(k) \end{aligned} \qquad (2)$$

where $w(k)$ is the residual interference introduced during the interference cancellation processing of our BKIC scheme. According to (2), we have the following two definitions:

**Definition 1**: Desired Signal (DS): $x'(k) = \sum_{l=0}^{L_x} h_x(k,l)x(k-l)$, which does not contain any noise or interference and it is the target of the whole system.

**Definition 2**: Desired Signal plus Noise (DSN): $x'(k)+n(k)=\sum_{l=0}^{L_x}h_x(k,l)x(k-l)+n(k)$, which is equivalent to the received signal from a pure point-to-point transmission without any interference. The function of BKIC is to remove the known interference. In the ideal case, its output should be exactly DSN. In reality, BKIC needs to estimate DSN as accurately as possible.

The signal $z(k)$ in (2) is DSN with a small extra noise, from which the traditional signal detection algorithm can then proceed to detect *x* as in conventional receivers. As will be shown later, the small noise $w(k)$ can be approximated as a Gaussian noise with negligible variance. Therefore, any transmitting specifications employed by source *S* or any channel environment experienced by *S* do not affect our BKIC scheme.

A relay enabled with our interference cancellation may be built as in the system architecture shown in Fig. 2. When the signal is received, the relay first checks if the known interference is present with the block "interference check". Interference check may be implemented with data sequence correlation [18, 19]. To limit the scope of this paper, we will not delve into the details of "interference check". If the interference is not present, then the packet is directly fed to the conventional receiver for data detection. If known interference is present, then the packet is fed to the BKIC block to cancel the known interference. After that, the signal is fed to the conventional receiver for target data detection.

### III. Blind Known Interference Cancellation in Flat Fading channel

In this section, we present our BKIC schemes assuming flat fading channel. The next section extends the treatment to the general multipath channel. BKIC consists of two steps. In the first step, the interference is canceled by combining adjacent symbols. In the second step, the DSN, i.e., the point-to-point form of the signal, distorted during the cancellation step, is recovered. The second step is the non-trivial step, and we present two algorithms for it. The first algorithm, which serves more like a benchmark, recovers DSN by means of smoothing; the second algorithm, by a novel real-valued belief propagation framework specially designed for our purpose here.

**Step 1: Interference cancellation**

With flat fading, $L_I=1$ for the interference channel, we can rewrite (1) as

$$r(k) = x'(k) + h_I(k)I(k) + n(k). \tag{3}$$

In practice, the channel may be time varying, and the variation depends on the moving speed and other environmental factors. However, for adjacent symbols, the channel variation is very small. Then we can approximate the channel variation as

$$h_I(k+1) \approx h_I(k) + \Delta(k). \tag{4}$$

In (4), $\Delta(k)$ is governed by the Doppler rate, and it is almost negligible in modern wireless communication systems [20, 24]. This is a key observation in this paper that enables us to use adjacent symbols to cancel the known interference without channel estimation. To do so, we obtain a new signal $t(k)$ by combining $r(k)$ and $r(k+1)$ as follows:

$$\begin{aligned} t(k) &= r(k) - \frac{I(k)}{I(k+1)} r(k+1) \\ &= \left( x'(k) - \frac{I(k)}{I(k+1)} x'(k+1) \right) + n(k) - \frac{I(k)}{I(k+1)} n(k+1) - I(k)\Delta(k) \end{aligned} \quad k \in \{1, 2, \cdots N-1\} \tag{5}$$

In (5), almost all the interference terms have been removed from $t(k)$. However, the signal of interest to us is DSN, $x'(k) + n(k)$, rather than $t(k)$. In the next step, we show how to extract DSN from $t(k)$.

**Step 2: Interested signal recovery**

In (5), the target signal $x'(k)$ is distorted into the form of $t(k)$ after the interference cancellation step. At first glance, $t(k)$ may appear to be the signal $x'(k)$ passing through an Inter-Symbol Interference (ISI) channel, in which case traditional ISI equalization schemes such as filtering, Viterbi detection and Belief Propagation (BP) [21] detection, could be used to recover $x'(k)$. However, a closer examination reveals an important difference between the signals $t(k)$ in (5) and that from a traditional ISI channel. Specifically, the difference is the correlated noise in (5) for adjacent symbols $t(k)$ and $t(k+1)$. Although Viterbi/BP detection achieves optimal MAP performance for independent noise in ISI equalization, its performance is far from optimal for the recovery of the target signal here because of the correlated noise, as will be shown in our numerical simulation.

Noise whitening is a standard technique for dealing with correlated noise. However, we cannot directly whiten the noise in (5) because it is impossible to transform the *N*-1 equations in (5) into *N* equations with independent noise terms while maintaining the interference cancellation effect. Another noise whitening scheme is noise prediction and whitening process in [22]. As will be shown in our simulation results later, this scheme is also far from optimal.

We now propose two schemes to recover DSN with near optimal performance.

*Recovery by Smoothing*

From (5), we could write

$$u(1) = t(1) = x'(1) + n(1) - \frac{I(1)}{I(2)}(x'(2) + n(2)) - I(1)\Delta(1)$$

$$u(k) = u(k-1) + \frac{I(1)}{I(k)}t(k) = x'(1) + n(1) - \frac{I(1)}{I(k+1)}(x'(k+1) + n(k+1)) - I(1)\sum_{m=1}^{k}\Delta(m) \quad (6)$$

$$\text{for } k \in \{2, \cdots N\text{-}2\}$$

Then, we obtain the estimate of $x'(1) + n(1)$ as follows:

$$z(1) = \frac{1}{N-1}\sum_{k=1}^{N-1}u(k) = x'(1) + n(1) + w(1) \quad (7)$$

where the residual interference $w(1) = -\frac{1}{N-1}\sum_{k=1}^{N-1}\left\{\frac{I(1)}{I(k+1)}(x'(k+1) + n(k+1)) + I(1)(N-k)\Delta(k)\right\}$ is independent of *x'*(1) and *n*(1). As will be shown in the next section, *w(k)* can be approximated by Gaussian distribution. For slow fading ($\alpha$ almost equal 1), its variance is very small; for fast fading, it may become larger due to the accumulation of errors (note: this effect is inevitable in any channel estimation scheme). Then, we can remove $x'(1) + n(1)$ from each signal in (5) to obtain the estimate of $x'(k) + n(k)$ as:

$$z(k) = \frac{I(k)}{I(1)}(z(1) - u(k-1)) = x'(k) + n(k) + I(k)(w(1)/I(1) + \sum_{m=1}^{k}\Delta(m)) = x'(k) + n(k) + w(k) \quad \text{for } k \geq 2 \quad (8)$$

We can then fed *z(k)* to a conventional receiver for final data detection.

*Recovery by Real-valued BP*

Belief propagation is a powerful technique to infer information from a large amount of correlated data. In the conventional method of applying BP to equalize ISI as in (5), ***x'*** and its ISI form

$x'(k) - \frac{I(k)}{I(k+1)}x'(k+1)$ are associated with the corresponding variable nodes. Their estimates are refined with message passing [21]. Direct BP application as such assumes the noise terms in the $N-1$ equations in (5) are independent. Strictly speaking, this is not true.

For correlated noise terms, [22] proposed to predict and whiten the correlated noise during the message passing procedure to improve the performance of the traditional BP algorithm. However, this method cannot make full use of the special noise correlation form in (5) and achieve only suboptimal performance. For better performance, we propose a novel BP scheme where DSN ***x'+n*** and the post-cancellation signal ***t***, rather than DS as in traditional BP, are associated with the variable nodes. Since they are real-valued signals, we refer to our BP detection algorithm as BKIC with Real-valued BP (BKIC-RBP). The Tanner graph of our BP algorithm is shown in Fig. 3.

An important subtlety in the Tanner graph is that ***x'+n*** is treated as "signal symbols". With respect to Fig. 2, ***x'+n*** is the target signal that will be fed to the conventional receiver after the interference cancelation process. With reference to (5), the observation $t(k)$ are made up of adjacent "signal symbols" plus noise in the cancellation process, which is $-I(k)\Delta(k)$ and does not include $n(k)$ and $n(k+1)$. The noise $n(k)$ will be dealt with by the conventional receiver later.

**Remark:** We stress that associating DSN rather than DS to the left variable node in Fig. 3 is the key of our BKIC-RBP scheme. First, directly estimating DS cannot improve the performance since the noise is independent of the interference. More importantly, $n(k) - \frac{I(k)}{I(k+1)}n(k+1) - I(k)\Delta(k)$ becomes the general noise in ***t*** and the relation between adjacent noise terms in (5) is wiped off from the figure when associating DS to the left variable node.

With the above setting, the target of BKIC-RBP is to find a vector $\boldsymbol{x'+n}$ to maximize

$$\begin{aligned}P(\boldsymbol{x'+n}|\boldsymbol{t}) &\propto P(\boldsymbol{t}|\boldsymbol{x'+n})P(\boldsymbol{x'+n}) \\ &= \prod_k P(t(k)|x'(k)+n(k), x'(k+1)+n(k+1))P(\boldsymbol{x'+n})\end{aligned} \quad (9)$$

Based on (9), a corresponding Tanner Graph can be established as in Fig. 3, where the messages being passed between the variable nodes and the check nodes are the probability density functions of the variable nodes at left hand side. The algorithm includes three critical steps: initializing the messages, updating the messages at the variable nodes and updating the messages at check nodes in an iterative way.

*Message initialization:*

The variable nodes ***x'+n*** on the left side of Fig. 3 are not associated with any channel outputs, and we initialize the PDF of ***x'+n*** with the *a priori* probabilities. The variable *x'* adopts a discrete value, and the discrete distribution is determined by the constellation set and the multipath channel. However, we cannot obtain the distribution because we assume channel information is not available in BKIC. Generally speaking, the upper bound of the interference power, $P_{max}$, can be derived easily. For example, $P_{max}$ could be set to the max power of the received signal ***r***[1].

We assume *x'* is uniformly distributed between $-\sqrt{P_{max}}$ and $\sqrt{P_{max}}$ [2]. Since the noise is of Gaussian distribution, the messages (i.e., the *a priori* probabilities) associated with the leftmost edges in the Tanner graph can be expressed as

$$m_{x'+n} = p_{x'+n}(y) = p_{x'+n}(x'+n=y) = \int_{-\sqrt{P_{max}}}^{\sqrt{P_{max}}} p_{x'}(x'=s) p_n(n=y-s) ds$$
$$= \frac{1}{2\sqrt{P_{max}}\sqrt{2\pi}\sigma} \int_{-\sqrt{P_{max}}}^{\sqrt{P_{max}}} \exp\left(-(y-s)^2/2\sigma^2\right) ds \quad . \tag{10}$$

For each left variable node $x'(k)+n(k)$, there is a incoming edge from an adjacent check node whose messages is denoted by $\bar{m}_{x'(k)+n(k)}$ and an outgoing edge to an adjacent check node whose message is denoted by $\vec{m}_{x'(k)+n(k)}$. The initial values of both $\bar{m}_{x'(k)+n(k)}$ and $\vec{m}_{x'(k)+n(k)}$ are set to $m_{x'+n}$.

With the initial messages, we then iteratively update them to obtain the final estimation. Since our Tanner graph in Fig. 3 does not include any circles, one iteration is enough to obtain the optimal MAP performance. Each iteration includes two parallel message update processes. One successively updates the messages, $\bar{m}_{x'(k)+n(k)}$ and $\vec{m}_{x'(k)+n(k)}$, one by one from top to bottom as illustrated in Fig. 3(a). The other process successively updates them from bottom to top as illustrated in Fig. 3(b). For example, for top-to-bottom message updates, the check-node update rule and the variable-node update rule in the following are applied in an alternating manner because each check-node message update depends on the previous variable-node message update, and vice versa.

---

[1] $P_{max}$ obtained in this way also includes the power of DSN and the power of the interference. So it is a loose upper bound of the maximal interference power.
[2] A more accurate distribution of *x'+n* should improve the performance of BKIC-RBP. Fortunately, BKIC-RBP with the approximate a prior distribution still performs very well, as demonstrated by our simulation results later.

*Message updates at the check nodes:*

First consider the top-to-bottom process where the messages associated with the edges between the left variable nodes and the check nodes are updated one by one from top to bottom as in Fig. 3 (a). For a check node connected to the right evidence node $t[k]$, the two left variable nodes connected to it are $x'(k)+n(k)$ and $x'(k+1)+n(k+1)$. Given the input PDF as $\vec{m}_{x'(k)+n(k)}$, then the output PDF can be calculated according to (5):

$$\begin{aligned}\vec{m}_{x'(k)+n(k)} = p_{x'(k+1)+n(k+1)}(y) &= \left|\frac{I(k+1)}{I(k)}\right| p_{x'(k)+n(k)}\left(\frac{I(k)}{I(k+1)}y+t(k)+I(k)\Delta(k)\right) \\ &= \left|\frac{I(k+1)}{I(k)}\right| \int_{\Delta(k)} p_{x'(k)+n(k)}\left(\frac{I(k)}{I(k+1)}y+t(k)+I(k)s\right) p(\Delta(k)=s)ds \\ &\propto \int p_{x'(k)+n(k)}\left(\frac{I(k)}{I(k+1)}y+t(k)+I(k)s\right) e^{-s^2/2\sigma_\Delta^2} ds\end{aligned} \quad (11)$$

where $\sigma_\Delta^2$ is the upper bound of the variance of the interference term $\Delta(k)$ [3]. For block fading, $\Delta(k)=0$ with probability 1. The PDF of $x'(k+1)+n(k+1)$ in (11) can be simplified to

$$\vec{m}_{x'(k)+n(k)} = p_{x'(k+1)+n(k+1)}(y) \propto p_{x'(k)+n(k)}\left(\frac{I(k)}{I(k+1)}y+t(k)\right). \quad (12)$$

Now consider the bottom-to-top process where the messages associated with the edges between the left variable nodes and the check nodes are updated one by one from the bottom to the top as in Fig. 3 (b). Analogous to (11), the PDF of $x'(k)+n(k)$ can be updated from the PDF of $x'(k+1)+n(k+1)$ and the observation $t(k)$ based on the following equation:

$$\vec{m}_{x'(k)+n(k)} = p_{x'(k)+n(k)}(y) \propto \int p_{x'(k+1)+n(k+1)}\left(\frac{I(k+1)}{I(k)}y-\frac{I(k+1)}{I(k)}t(k)-I(k+1)s\right) e^{-s^2/2\sigma_\Delta^2} ds \quad (13)$$

For block fading, (13) can be simplified to

$$\vec{m}_{x'(k)+n(k)} = p_{x'(k)+n(k)}(y) \propto p_{x'(k+1)+n(k+1)}\left(\frac{I(k+1)}{I(k)}y-\frac{I(k+1)}{I(k)}t(k)\right) \quad (14)$$

*Message updates at the variable nodes:*

---
[3] BKIC-RBP is robust to $\sigma_\Delta^2$ as shown in the simulation. Therefore, we can fix it to a relative high value, such as 0.001 with almost no performance loss.

Message updates at the left variable nodes is the same for both top-to-bottom process and the bottom-to-top process. Each left variable node is connected to three edges, whose associate messages are output PDF $\vec{p}_{x'(k)+n(k)}$, input PDF $\bar{p}_{x'(k)+n(k)}$ and the *a priori* PDF $p_{x'(k)+n(k)}$ respectively. Each output PDF is updated with the input PDF and the *a priori* PDF as

$$\vec{p}_{x'(k)+n(k)}(y) = \frac{1}{T}\bar{p}_{x'(k)+n(k)}(y) \cdot p_{x'(k)+n(k)}(y). \tag{15}$$

where *T* is the normalization factor.

At the end of the processing, we need to collect the information contained in all the messages and make a final estimate of DSN. For the *k*-th variable node with $x'+n$, there is the *a priori* PDF $p_{x'(k)+n(k)}(y)$, the input PDF $\bar{p}_{x'(k)+n(k)}(y)$ after the top-to-bottom process, and the input PDF $\bar{p}^*_{x'(k)+n(k)}(y)$ after the bottom-to-top process (there is only one input PDF of the first and the last left variable node). Then the final probability distribution of $x'(k)+n(k)$ can be calculated as the following product:

$$p^f_{x'(k)+n(k)}(y) = \frac{1}{T}\bar{p}^*_{x'(k)+n(k)}(y)\bar{p}_{x'(k)+n(k)}(y)p_{x'(k)+n(k)}(y) \tag{16}$$

where *T* is the normalization factor. $p^f_{x'(k)+n(k)}(y)$ contains all the information about *x'+n* and it can be fed to the traditional detection block for further target data detection. In order to compare with the BKIC-S scheme, an estimate of $x'+n$ is given by

$$z(k) = \arg\max_y p^f_{x'(k)+n(k)}(y) \tag{17}$$

***Discussion:*** we can regard *t* as an inner encoder output with input *x'*. Then the decoder of it, i.e., the hard/soft decision based on $p^f_{x'(k)+n(k)}(y)$ can be combined with the channel decoding procedure so that the Turbo like detection-decoding can applied to achieve even better performance.

**IV. BKIC in Frequency Selective Fading channel**

The previous section proposes the BKIC scheme for flat fading channel. In this section, we extend the scheme to the frequency-selective channel. When BKIC is performed in a totally blind manner, we have no prior information about the multi-path characteristics. In this case, it is still reasonable to assume knowledge of the maximum delay of all the paths, *L*, as in many current broadband wireless systems. For example, the length of the predefined CP (cyclic prefix) in OFDM system implies the maximum delay of all paths. Therefore, we can perform interference cancellation and recover the equivalent DSN for each potential path in a successive way as follows.

We first rewrite the received signal in (1) as

$$r(k) = x'(k) + \sum_{d=0}^{L} h_I(k,d)I(k-d) + n(k)$$
$$= x'(k) + \sum_{d=1}^{L} h_I(k,d)I(k-d) + h_I(k,0)I(k) + n(k) . \quad (18)$$
$$= x_0'(k) + h_I(k,0)I(k) + n(k)$$

In (18), we select the first path of the interfering signal as the interference to be cancelled and combine the other interfering paths to DS, $x'(k)$. Then, the selected interfering path can be removed by applying the BKIC scheme as in the preceding section. After that, we can obtain

$$z_1(k) = x'(k) + \sum_{d=1}^{L} h_I(k,d)I(k-d) + w_0(k) + n(k) \quad (19)$$

where $w_0(k)$ is the residual interference after removing the first path.

Comparing $z_1(k)$ and $r(k)$, we see that the first path of interference has been removed and generated a new residual interference term $w_0$. In a similar way, we can remove the second path of interference to obtain $z_2(k)$. Repeat the BKIC scheme $L+1$ times, we finally obtain that

$$z(k) = z_{L+1}(k) = x'(k) + \sum_{d=0}^{L} w_d + n(k) . \quad (20)$$

## V. Performance Analysis

In this section, we analyze the BER and SINR performance of the proposed BKIC schemes under flat fading channels and frequency selective channels.

**1. BKIC under Flat Fading Channel**

***Proposition 1:*** The performance of the BKIC schemes is upper bounded by the clean system where there is no interference at all.

This proposition is easy to understand since the interfering data is independent to the target data and it can not help to detect the target data. Due to the existence of the residual interference $w$, BKIC schemes can never achieve this upper bound exactly. Fortunately, we can approach it very closely.

***Proposition 2:*** For the BKIC-S scheme, the residual interference $w(k)$ can be well approximated by

a Gaussian noise $N(0,\mu(k))$, where the variance is upper bounded by

$$\mu(k) = I^2(k)\left(\frac{(P_x+\sigma^2)}{(N-1)} + \frac{NP_I}{3}(1-\alpha^2)\right)E\{1/I^2(j)\}.$$

Proof: According to (8), we can show that the residual interference has the largest variance for the first symbol, which can be expressed as $w(1) = -\frac{I(k)}{N-1}\sum_{j=1}^{N-1}\left\{\frac{1}{I(j+1)}(x'(j+1)+n(j+1))-(N-j)\Delta(j)\right\}$. When $N$ is very large, the total residual interference, $w$, can be regarded as of normal distribution based on large number theory. It is straightforward to verify that its mean value is zero and its variance is

$$\begin{aligned}\mu(k) \leq \mu(1) &= E\{|w(k)|^2\} = \frac{(P_x+\sigma^2)}{(N-1)}E_j\{\frac{I^2(k)}{I^2(j)}\} + I^2(k)\frac{N(2N-1)}{6(N-1)}\sigma_\Delta^2 \\ &\approx I^2(k)\left(\frac{(P_x+\sigma^2)}{(N-1)} + \frac{NP_I}{3}(1-\alpha^2)\right)E\{1/I^2(j)\}\end{aligned} \quad (21)$$

where $P_x$ is the received power of the target signal $x'$ and $P_I$ is the received power of the interfering signal.

From the above proposition, we can obtain some important observations.

***Corollary 1:*** In BKIC-S, the residual interference is independent of the received power of the interfering signal for block fading channel.

In block fading channel, $\alpha=1$ and $\sigma_\Delta^2=0$. Then the power of the interfering signal, embedded in $h_I$, does not affect the performance of the BKIC-S scheme. In general slow fading channel, the block fading assumption holds very well. This independent property is desired especially when the interfering signal power is much stronger. In contrast, the traditional known interference cancellation [3] scheme, which is based on estimated channel information, performs poorly in this case because the residual interference is proportional to the interfering signal power with a given channel estimation mean square error (MSE). In block fading, the residual interference is very small for large packet length. For example, when $N$=100, the residual interference power is decreased by about -20dB. Assuming equal interference power and DS power, the residual interference is 20dB less than the DS power, which is much smaller than the general 10dB SINR requirement for wireless receiver.

***Corollary 2:*** In BKIC-S, there is an optimal packet length[4] $N$ to minimize the residual interference $\mu$ for continuous fading channel.

The first term in (21) decreases with $N$ as in the block fading channel. The second term in (21) increases with $N$. When $N$ is large, the channel varies far from its average value and the residual interference coming from channel variation accumulates (fortunately, BKIC-RBP is insensitive to channel variation.). If the interference data adopts a constant power modulation (PSK modulation), the optimal $N$ can be calculated as

$$\partial \mu / \partial N = 0 \quad \Rightarrow \quad N_{opt} = 1 + \sqrt{\frac{3(P_x + \sigma^2)}{\sigma_\Delta^2}} \tag{22}$$

In real communication systems, $\sigma_\Delta^2$ is very small and $N_{opt}$ is large.

Besides the performance, complexity is also an important issue. The complexity of our BKIC-S scheme is quite low. Only one multiplication and two addition processes are needed for each symbol.

***Corollary 3***: With BKIC-S, the SNR loss compared to the clean system (only DSN signal exists) is

$$\Delta = SNR_{TSN} - SNR_{BKIC-S} = 10\log(\frac{P_x}{\sigma^2}) - 10\log(\frac{P_x}{\sigma^2 + \mu}) = 10\log(1 + \frac{\mu}{\sigma^2}) \approx \mu SNR_{TSN} \quad . \tag{23}$$

In BKIC-S, the residual interference is fixed when the power of DS plus noise is given. As a result, the SNR loss of BKIC-S depends on the SNR of DSN. For smaller residual interference, the SNR loss is approximately proportional to the SNR of DSN.

For the performance of BKIC-RBP, we have

***Proposition 3***: The performance of the BKIC-RBP scheme achieves the MAP optimal DSN recovery performance, which is lower bounded by the BKIC-S.

As is well known, the loop free BP detection has MAP-optimal performance. In our BKIC-RBP, there are no circles in the Tanner graph in Fig. 3, so exact MAP performance of signal recovery can be achieved. As a result, the performance of BKIC-RBP is lower bounded by the BKIC-S scheme.

***Proposition 4:*** The complexity of BKIC-RBP is linear in terms of packet length.

---

[4] In our paper, packet length is just the processing length of the algorithms. Dividing a packet into several parts for processing is not considered.

The loop-free property in Fig. 3 also guarantees fast convergence of the BP algorithm. Only one iteration (one top-to-bottom process and one bottom-to-top process) in the RBP algorithm is enough to obtain the MAP performance. The complexity of our RBP recovery is only $4N-6$ message update operations, which is linear in terms of the packet length.

However, real valued processing is needed to achieve the optimal MAP performance. In practice, we need to quantize the real valued PDF into discrete form with controlled complexity. As shown in our simulation and many other works [23], there is typically little performance loss associated with quantization errors in belief propagation algorithms.

**2. Analysis of selective fading BKIC**

In multipath channel, the flat fading BKIC is executed several times to successively get rid of the interference of each path. Therefore, the performance analysis is similar. We have

*Proposition 5:* For the BKIC-S in multipath channel, the total residual interference $w$ can be approximated by a Gaussian noise $N(0, \sum_{m=0}^{L} \mu_m)$, where $\mu_m$ is the variance of the residual interference generated after cancellation of the $m$-th path. Its value can be obtained as in (21).

According to *Proposition 5*, we can obtain the corresponding corollaries as those in the flat fading channel.

With respect to *Proposition 3*, we can obtain a similar proposition as follows:

*Proposition 6*: In the multipath channel, BKIC-RBP achieves the MAP optimal performance for each path. Its performance is lower bounded by the performance of BKIC-S.

**VI. Numerical Simulation**

This section presents numerical simulation results for the performance of BKIC. Without loss of generality, we assume the BPSK modulation, i.e., $x, I \in \{1, -1\}$, for both target signal and the interfering signal. As is clear from the earlier discussion, the target signal does not affect the operation of BKIC. Thus, without loss of generality in BKIC, we could assume flat fading with unit channel coefficient for the target-signal channel (note: with respect to Fig. 2, it is in the conventional receiver that the channel characteristics of the target signal that matters, and it is over there that the actual fading characteristics of the target signal channel come into play).

For the interference channel, block fading, continuous fading channel and multi-path fading are simulated. The system SNR is defined as $1/\sigma^2$, where 1 is the power of the target signal. For BKIC-RBP, we need to quantize the messages (the PDFs) into discrete form to enable simulation with Matlab. In our simulation experiments, the quantization interval for BKIC-RBP is 0.025 for 1-7dB and it is 0.0125 for 8-10dB. Quantization of the messages results in quantization errors, and smaller quantization step can further improve the performance at the cost of high complexity.

For comparison purposes, we also simulate the scheme of Noise Predictive Belief Propagation (NPBP) in [22] for the second step of BKIC. In our simulation, the length of the whitening filter in [22] is set to two. The complexity of BKIC-NPBP is $30(4N-6)$ message update operations plus 3 whitening operations, which is much higher than BKIC-S and BKIC-RBP.

***Block fading:***

We first consider single path block fading interference channel. In this case, the channel coefficients are set to a constant unit within the whole block.

In Fig.4 and Fig. 5, we show the residual interference variance and BER of the proposed BKIC schemes respectively, with packet length *N*=100 (bits).

We first look at the simulation performance of the BKIC-S. For BKIC-S, the theoretical variance of the residual interference is $\mu = \frac{(1+\sigma^2)}{(N-1)}$ according to (21). Based on it, we can calculate the theoretical BER of BKIC-S as $Q(\sqrt{1/(\mu+\sigma^2)})$ as in [24]. In both figures, the simulation results match the theoretical values very well. This validates the Gaussian approximation in *Proposition 2*. In Fig. 4, we can see that the residual interference of BKIC-S is almost independent of the SNR. The near-constant residual interference becomes more significant compared to noise in high SNR region. As a result, compared to the lower bound (i.e., the standard BPSK without interference), the SNR loss in Fig. 5 is larger when SNR increases. The SNR at BER of 2E-3 is around 10dB, corresponding to an SNR loss of than 0.5 dB.

As predicted by *Proposition 3*, the performance of BKIC-RBP, including residual interference and BER, is better than BKIC-S. The improvement becomes even larger in high SNR region. Both figures show that BKIC-RBP can benefit more from the SNR increase. The reason is that high SNR will sharpen *a priori* distribution of DSN, which improves the performance of BKIC-RBP. By contrast, the conventional NPBP scheme is worse than BKIC-RBP by about 1.5dB. If we do not know the channel is block fading or continuous fading, and fixed $\sigma_\Delta^2$ as 0.001, the performance of BKIC-RBP1 is obtained. The almost identical performance shows that BKIC-RBP is robust to $\sigma_\Delta^2$.

Fig. 6 shows the BER performances of different schemes for $N$=1000. As predicted in the analysis, the BER performance of BKIC-S is significantly improved as $N$ increases. It is almost the same as the theoretical lower bound. On the other hand, the BER performance improvement of BKIC-RBP is negligible compared to the case where $N$=100 due to the fixed quantization error in our simulation.

*Continuous Fading:*

In this part, we set the interference channel to single path continuous fading with the first order Markov channel mode in as in [17] and the parameter $\alpha$ is set to $1-10^{-3}$, which corresponds to a fast fading channel. $\sigma_\Delta^2$ is set to 0.001.

Fig. 7, and Fig. 8 show the residual interference variance and BER of the BKIC schemes when $N$=100, respectively. For comparison, we also give the BER performance of the traditional known interference cancellation (Traditional KIC) scheme where the channel coefficient of the first symbol is perfectly known. Therefore, the given performance is an upper bound of the actual Traditional KIC with non-ideal channel estimation. From both figures, BKIC-RBP outperforms all the other schemes by at least 1dB. BKIC-S is next to the best scheme when SNR is less than 10 dB. Compared to Fig. 5, all schemes degrade in continuous fading channel. The degradation for BKIC-RBP and BKIC-NPBP is about 0.1 dB, which is much smaller than that of BKIC-S and traditional KIC.

In Fig. 9, we present the BER performance of the BKIC schemes when $N$=1000. With large $N$, the channel varies more significant and the BKIC-S scheme performs even worse. The BKIC-RBP scheme performs best among all the schemes and there is only 0.2 dB SNR loss compared to the block fading case. BKIC-RBP is about 1dB better than the BKIC-NPBP scheme. It indicates that the BP algorithm is not significantly affected by channel variation.

*Frequency Selective Fading:*

We also investigate the BER performance of BKIC with frequency selective fading channel. The results are shown in Fig. 10. For illustration, a simple multipath interference channel scenario where there are two interference paths is considered. The amplitudes of the two paths are the same and unchanged within the packet; the delays of the two paths are 0 and 2 respectively. Compared to the flat fading case in Fig.5, both the performances of BKIC-RBP and BKIC-S are degraded as analyzed in the previous section. BKIC-S degrades more than BKIC-RBP, which has only tiny degradation.

## VII. Conclusions

This paper presents two known interference-cancellation schemes with good performance and low complexity. Although there has been much theoretical work in this area, deployments of the previously proposed schemes are difficult because of their needs for accurate channel estimation and their high complexity. To our knowledge, there has been no effective blind known interference-cancellation scheme that does not require estimation of the interference channel. Our work fills a gap in that regard.

Specifically, this paper proposes a framework for blind known interference cancellation (BKIC), as embodied in Fig. 2. The principle on which BKIC operates is based on the observation the channel coefficient is almost constant for adjacent interference symbols. Thus, if the interference symbols are known, by combining adjacent symbols (i.e., combining the received signal and a weighted off-shifted version of it), we can obtain a new signal that is almost free of the interference. This, however, causes distortion to our target signal. A key component of BKIC, therefore, is how to compensate for this distortion. To do so, we propose and investigate two schemes: BKIC-S, which is based on the principle of smoothing; and BKIC-RBP, which is based on the principle of real-valued belief propagation. BKIC-RBP has MAP-optimal performance. The algorithmic complexities of both schemes are linear in the size of the packet.

We show that both BKIC-S and BKIC-RBP have superior performance than the traditional schemes and their performance is very close to the theoretical performance bound, especially for block fading interference channel. The performance of BKIC-S improves with packet size, while the performance of BKIC-RBP is not sensitive to the packet size, but is dependent on the quantization step used in the algorithm to approximate real values. Importantly, BKIC-RBP is very robust against fast fading in which the channel coefficients may vary in a dynamic manner within a packet. Going forward, to fully exploit the potential of BKIC in wireless networks, new MAC layer and network layer protocols need to be designed. Besides relay networks, recently there has been increased interest in the wireless networking community on the realization of wireless full-duplex communication [25]. In the full duplex mode, a node transmits and receives at the same time. The received signal contains both the target signal as well as known interference (i.e., the self transmitted signal that is known). The investigation of BKIC for wireless full-duplex communication will be of much interest.

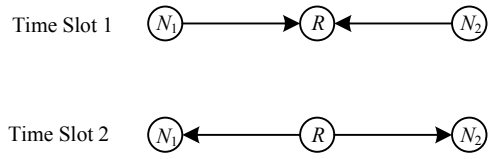

(a) Two way relay with analog network coding, where the two sources N1 and N2 transmit simultaneously to the relay in the first time slot, and the relay amplifies and broadcasts the received signal to both sources in the second time slot.

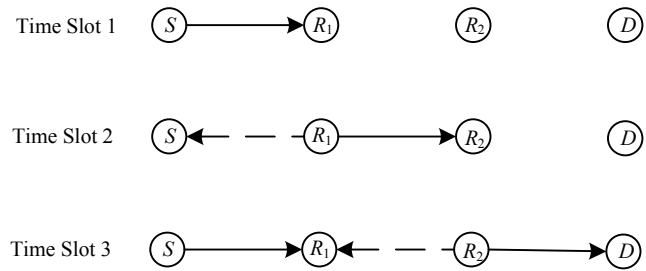

(b) One way relay channel in a chain.

Fig. 1. Two wireless networks with known interference.

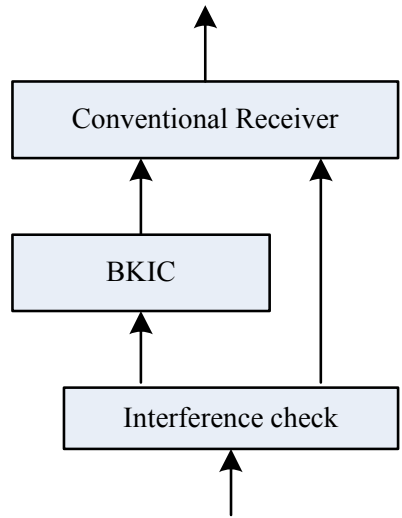

Fig. 2. System architecture with blind known interference cancellation

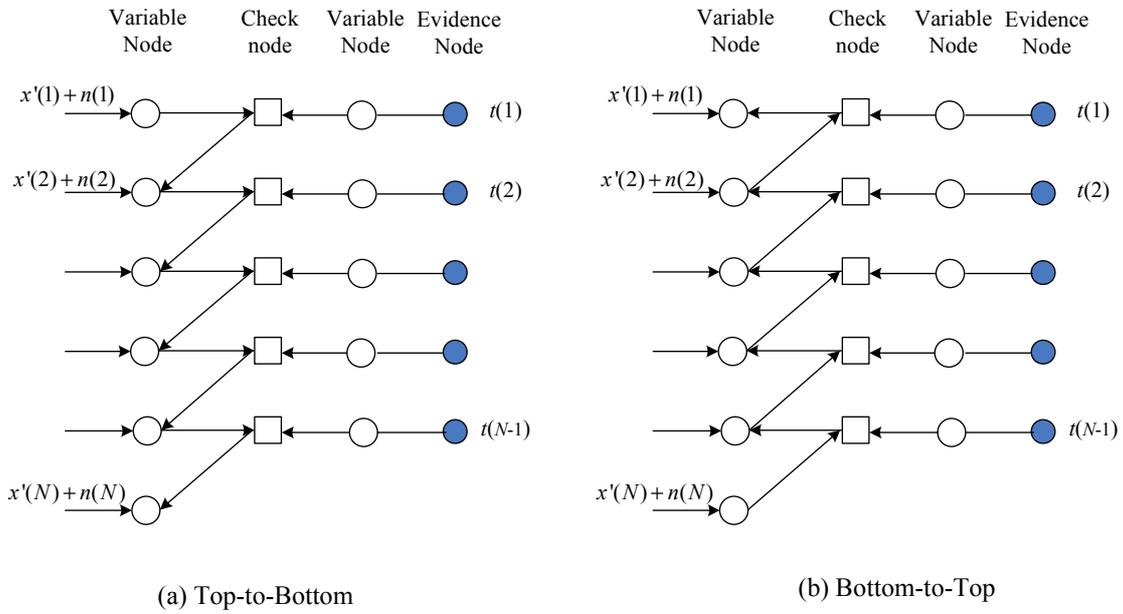

Fig. 3. Tanner Graph for Continuous BP, where blank circles denote the variable nodes and the filled circles denote the evidence nodes and the rectangles denote the check nodes

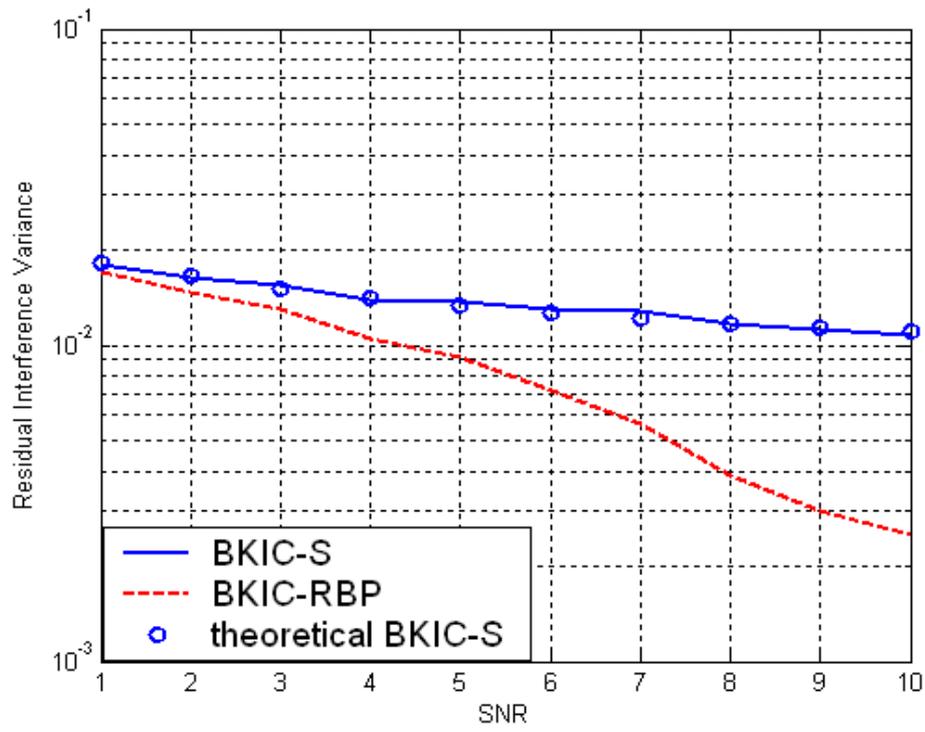

Fig. 4. Residual interference variance with $N$=100.

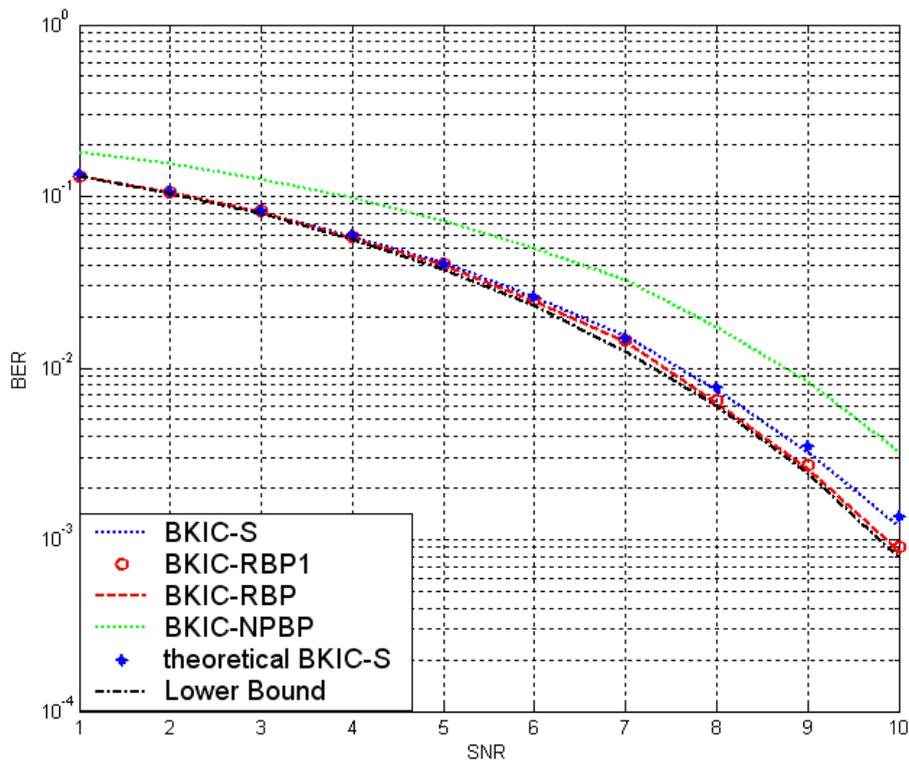

Fig. 5. BER performance for block fading with $N$=100.

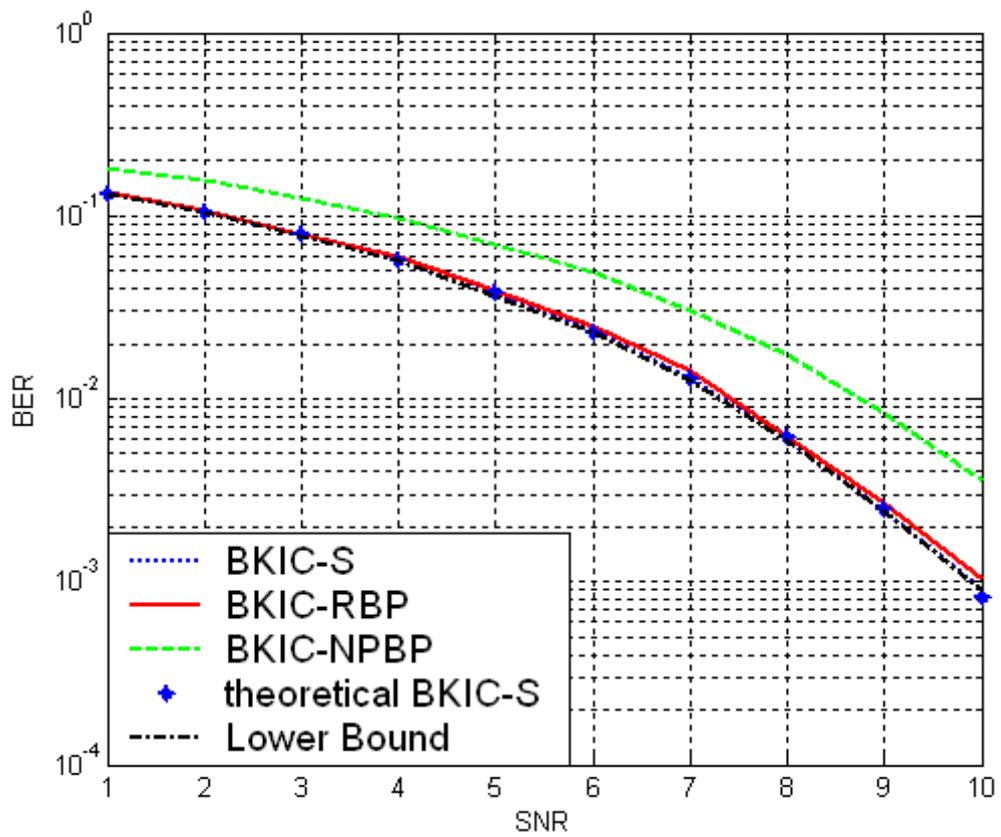

Fig. 6. BER performance for block fading with $N$=1000

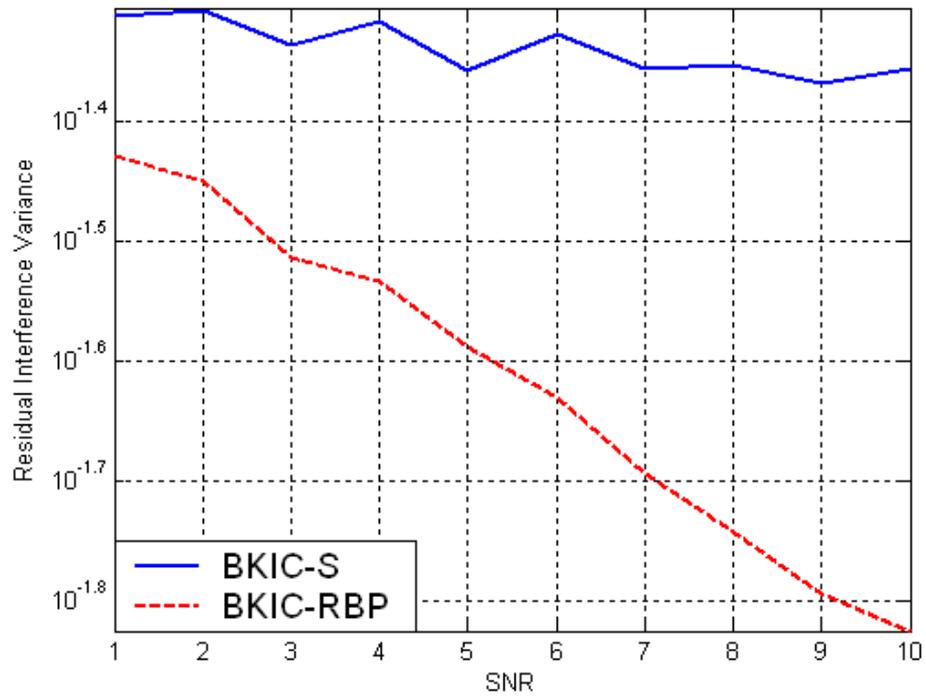

Fig. 7. Residual interference for continuous fading channel with *N*=100

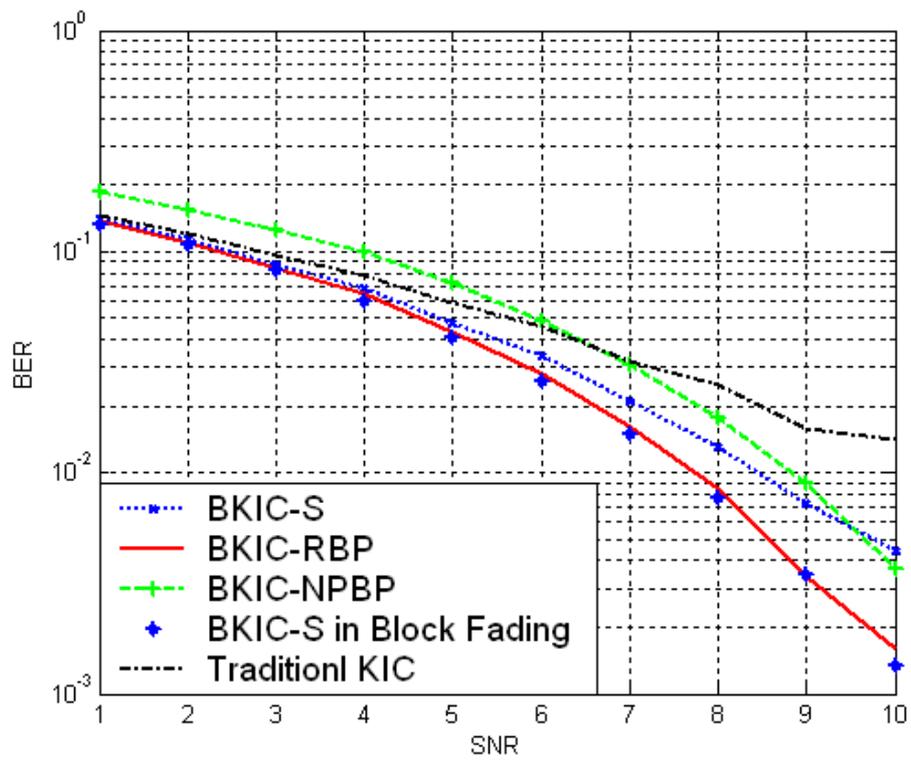

Fig. 8. BER performance for continuous fading channel with $N$=100.

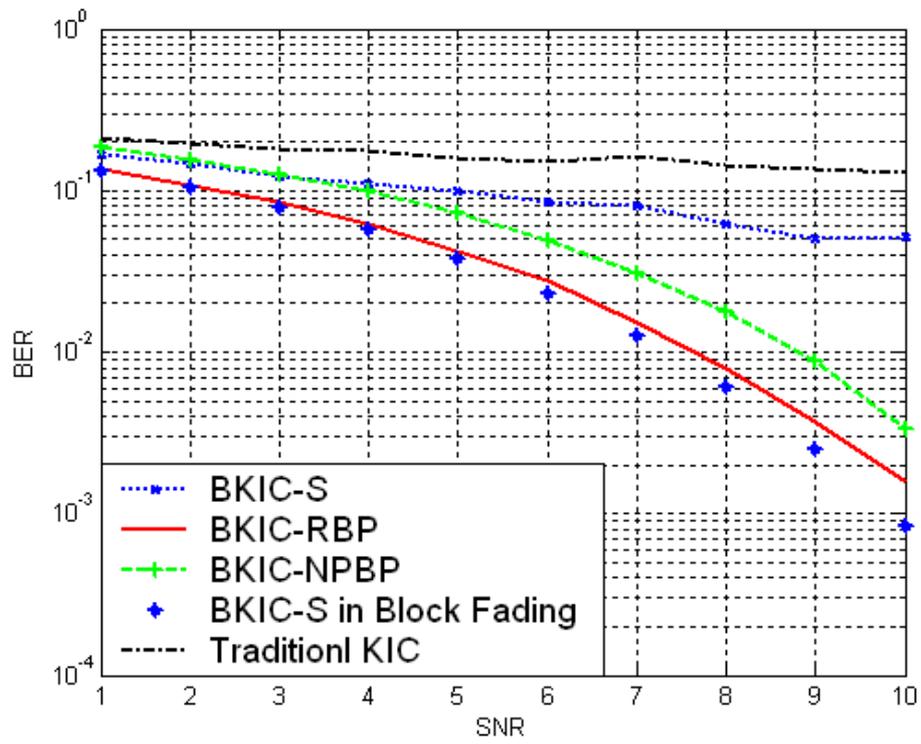

Fig. 9. BER performance for continuous fading channel with $N$=1000

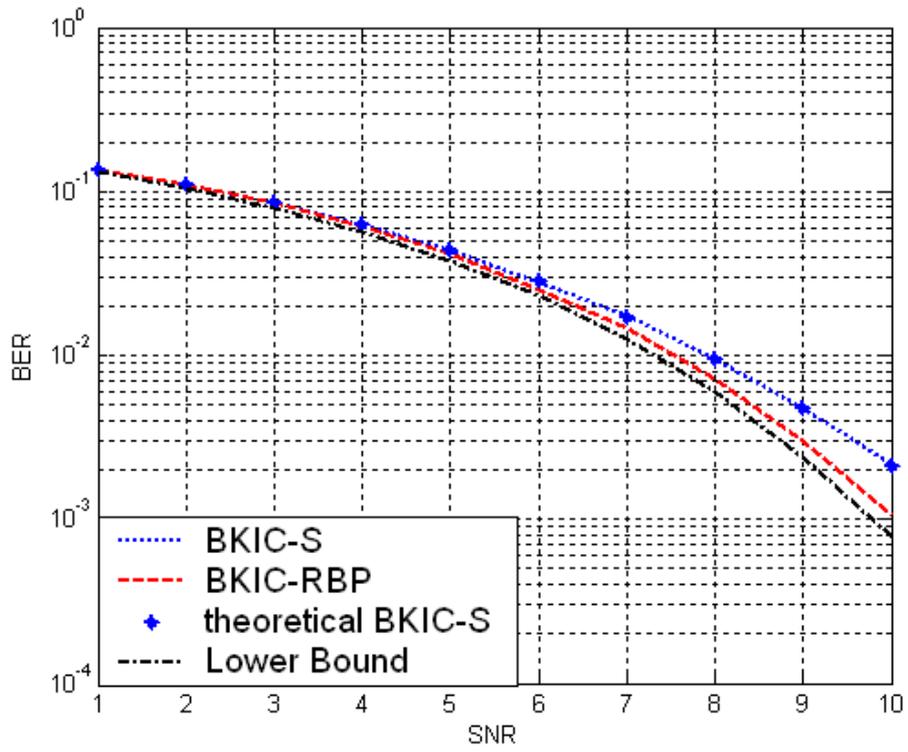

Fig. 10. BER performance for frequency selective fading channel with *N*=100